\documentclass[a4paper,11pt]{article}
\usepackage{amssymb}
\usepackage{latexsym,bm}

\def \be {\begin{equation}}
\def \ee {\end{equation}}
\def \bea {\begin{eqnarray}}
\def \eea {\end{eqnarray}}
\def \nn {\nonumber}

\def \a {\alpha}
\def \b {\beta}

\def \d {\delta}

\def \m {\mu}
\def \n {\nu}
\def \k {\kappa}

\def \s {\sigma}
\def \r {\rho}
\def \o {\omega}

\def \th {\theta}
\def \Th {\Theta}

\def \t {\tau}
\def \dag {\dagger}
\def \p {\partial}

\def\bd{\begin{document}}
\def\ed{\end{document}}
\def\nn{\nonumber}
\def\bea{\begin{eqnarray}}
\def\eea{\end{eqnarray}}
\let\bm=\bibitem
\let\la=\label

\def\N{{\cal N}}
\def\sst{\scriptscriptstyle}
\def\thetabar{\bar\theta}
\def\Tr{{\rm Tr}}
\def\one{\mbox{1 \kern-.59em {\rm l}}}

\def\a{\alpha}      \def\da{{\dot\alpha}}
\def\b{\beta}       \def\db{{\dot\beta}}
\def\c{\gamma}  \def\C{\Gamma}  \def\cdt{\dot\gamma}
\def\d{\delta}  \def\D{\Delta}  \def\ddt{\dot\delta}
\def\e{\epsilon}        \def\vare{\varepsilon}
\def\f{\phi}    \def\F{\Phi}    \def\vvf{\f}
\def\h{\eta}
\def\k{\kappa}
\def\l{\lambda} \def\L{\Lambda}
\def\m{\mu} \def\n{\nu}
\def\o{\omega}
\def\P{\Pi}
\def\r{\rho}
\def\s{\sigma}  \def\S{\Sigma}
\def\t{\tau}
\def\th{\theta} \def\Th{\Theta} \def\vth{\vartheta}
\def\X{\Xeta}
\def\z{\zeta}
\def\w{\wedge}
\def\u{\underline}
\def\hs{\hspace}

\def\cA{{\cal A}} \def\cB{{\cal B}} \def\cC{{\cal C}}
\def\cD{{\cal D}} \def\cE{{\cal E}} \def\cF{{\cal F}}
\def\cG{{\cal G}} \def\cH{{\cal H}} \def\cI{{\cal I}}
\def\cJ{{\cal J}} \def\cK{{\cal K}} \def\cL{{\cal L}}
\def\cM{{\cal M}} \def\cN{{\cal N}} \def\cO{{\cal O}}
\def\cP{{\cal P}} \def\cQ{{\cal Q}} \def\cR{{\cal R}}
\def\cS{{\cal S}} \def\cT{{\cal T}} \def\cU{{\cal U}}
\def\cV{{\cal V}} \def\cW{{\cal W}} \def\cX{{\cal X}}
\def\cY{{\cal Y}} \def\cZ{{\cal Z}}

\def\ua{\underline{\alpha}} \def\ubb{\underline{\beta}}
\def\ug{\underline{\gamma}}
\def\ub{\underline{\phantom{\alpha}}\!\!\!\beta}
\def\uc{\underline{\phantom{\alpha}}\!\!\!\gamma}
\def\um{\underline{\mu}} \def\un{\underline{\nu}}
\def\ud{\underline\delta}
\def\ue{\underline\epsilon}
\def\una{\underline a}\def\unA{\underline A}
\def\unb{\underline b}\def\unB{\underline B}
\def\unc{\underline c}\def\unC{\underline C}
\def\und{\underline d}\def\unD{\underline D}
\def\une{\underline e}\def\unE{\underline E}
\def\unf{\underline{\phantom{e}}\!\!\!\! f}\def\unF{\underline F}
\def\unm{\underline m}\def\unM{\underline M}
\def\unn{\underline n}\def\unN{\underline N}
\def\unp{\underline{\phantom{a}}\!\!\! p}\def\unP{\underline P}
\def\unq{\underline{\phantom{a}}\!\!\! q}
\def\unQ{\underline{\phantom{A}}\!\!\!\! Q}
\def\unH{\underline{H}}
\def\ul{\underline}

\def\As {{A \hspace{-6.4pt} \slash}\;}
\def\bs {{b \hspace{-6.4pt} \slash}\;}
\def\Ds {{D \hspace{-6.4pt} \slash}\;}
\def\ds {{\del \hspace{-6.4pt} \slash}\;}
\def\ss {{\s \hspace{-6.4pt} \slash}\;}
\def\ks {{ k \hspace{-6.4pt} \slash}\;}
\def\ps {{p \hspace{-6.4pt} \slash}\;}
\def\pas {{{p_1} \hspace{-6.4pt} \slash}\;}
\def\pbs {{{p_2} \hspace{-6.4pt} \slash}\;}

\def\Fh{\hat{F}}
\def\Vh{\hat{V}}
\def\Xh{\hat{X}}
\def\ah{\hat{a}}
\def\xh{\hat{x}}
\def\yh{\hat{y}}
\def\ph{\hat{p}}
\def\xih{\hat{\xi}}

\def\psit{\tilde{\psi}}
\def\Psit{\tilde{\Psi}}
\def\tht{\tilde{\th}}

\def\At{\tilde{A}}
\def\Qt{\tilde{Q}}
\def\Rt{\tilde{R}}
\def\Nt{\tilde{N}}

\def\at{\tilde{a}}
\def\st{\tilde{s}}
\def\ft{\tilde{f}}
\def\pt{\tilde{p}}
\def\qt{\tilde{q}}
\def\vt{\tilde{v}}
\def\nt{\tilde{n}}

\def\delb{\bar{\partial}}
\def\bz{\bar{z}}
\def\bD{\bar{D}}
\def\bB{\bar{B}}

\def\bk{{\bf k}}
\def\bl{{\bf l}}
\def\bp{{\bf p}}
\def\bq{{\bf q}}
\def\br{{\bf r}}
\def\bx{{\bf x}}
\def\by{{\bf y}}
\def\bR{{\bf R}}
\def\bV{{\bf V}}

\def\d{\delta}\def\D{\Delta}\def\ddt{\dot\delta}

\def\p{\partial} \def\del{\partial}
\def\xx{\times}
\def\uno{\mbox{1 \kern-.59em {\rm l}}}

\def\trp{^{\top}}
\def\inv{^{-1}}
\def\dag{{^{\dagger}}}

\def\pr{\prime}
\def\rar{\rightarrow}
\def\lar{\leftarrow}
\def\lrar{\leftrightarrow}

\begin{document}
\title{High Spin Topologically Massive   Gravity}
\author{Bin Chen$^{1,2,3}$\footnote{Email:bchen01@pku.edu.cn}\hs{2ex}Jiang Long$^1$\footnote{Email:lj301@pku.edu.cn}\hs{2ex}\\
\small{$^1$Department of Physics,
and State Key Laboratory of
Nuclear Physics and Technology,}\\
\small{Peking University, Beijing 100871, P.R. China}\\
\small{$^2$Center for High Energy Physics,
Peking University, Beijing 100871, P.R. China}\\
\small{$^3$ Kavli Institute for Theoretical Physics China,
CAS, Beijing 100190, P.R. China}\\
}
\date{\today}
\maketitle


\begin{abstract}
We propose an action to describe high spin topologically massive  gravity with a negative cosmological constant. With the frame-like fields and spin connections being combined into two gauge fields, the action includes two gauge field Chern-Simons actions with different levels, and also a linear term proportional to the difference of the gauge field strengths. Such linear term play the role of imposing the torsion free conditions for high spin fields. We discuss the gauge symmetry of this action and study the fluctuations around the AdS$_3$ vacuum. We show how to relate the the fluctuations of the gauge field to the Frondal fields, using the gauge symmetry. For the gauge group $SL(n,R)\times SL(n,R)$, we find that the fluctuations of all high spin fields up to spin $n$ satisfy third order differential equations, and  hence there generically exist  massive traceless and trace mode for every spin.
 \end{abstract}

\newpage






\section{Introduction}

In the past decade, the theory of high spin fields has drawn much attention, especially in AdS/CFT correspondence. It has been known for a long time that even though the free high spin fields could be defined in flat and curved spacetime\cite{Fronsdal:1978}, the interacting high spin theory could only be defined in Anti-de Sitter (AdS) and de Sitter (dS) spacetime\cite{Aragone:1979hx,Vasiliev:1980as, Vasiliev:1986td}. Especially in $D\geq 4$ spacetime, once an interacting higher spin field  is considered, all the spin fields up to spin $\infty$ have to be taken into account, with additional compensator fields. Nevertheless, it has been conjectured that Vasiliev's minimal bosonic high spin field theory in AdS$_4$ is holographically dual to the singlet sector of three-dimensional $O(N)$ model in the large N and free field limit\cite{KP}. The nontrivial checks on the three-point correlation functions have been done few years ago\cite{Yin1}.

Since the summer of last year, the study of the high spin AdS$_3$ gravity has been developing very quickly. In this case, the compensator fields are not needed and more importantly, it is possible to define an interacting high spin gravity theory with finite spin truncation\cite{Henneaux:2010xg,Theisen:2010}. In a remarkable paper\cite{Theisen:2010}, the authors showed that the high spin AdS$_3$ gravity theory with spin up to $n$ could be written as a Chern-Simons gravity theory with
$SL(n,R)\times SL(n,R)$ gauge group\footnote{For original study on the relation between AdS$_3$ gravity and Chern-Simons theory, see \cite{Achucarro:1987vz,Witten:1988hc}. For the earlier study of the high spin field theory in AdS$_3$ in the framework of Chern-Simons gravity, see \cite{Blencowe:1988gj,{Bergshoeff:1989ns}}.}. By imposing appropriate boundary condition and gauge choice, they showed that the asymptotic symmetry group of high spin AdS$_3$ gravity is a classical $W_n$ symmetry with the same central charge in the pure AdS$_3$ gravity found in \cite{Brown:1986nw}. This indicates that the high spin AdS$_3$ gravity is holographically dual to a conformal field theory with $W_n$ symmetry. Similar observation has been made in \cite{Henneaux:2010xg} starting from the $n\to \infty$ theory. For the further developments in high spin AdS$_3$ gravity and its dual, see \cite{Gaberdiel:2010ar,Gaberdiel:2010pz,Gaberdiel:2011wb, Gaberdielnew,{Castro:2010ce},Ahn:2011pv,{Chang:2011mz},Polyakov:2011sm,Campoleoni:2011hg,Castro:2011ui, Bergshoeff:2011pm,Lu:2011qx}. For the study of black holes with high spin hair, see \cite{Gutperle:2011kf,Ammon:2011nk,{Kraus:2011ds},Castro:2011fm}.

For pure gravity and high spin fields in three dimensions, there is no local degree of freedom, even though there do exist boundary degrees of freedom at asymptotic infinity, as shown in the existence of BTZ black hole and the black holes with high spin hair. One possible way to induce local dynamical degree of freedom without bringing in ghost is to introduce the gravitational Chern-Simons term and its cousin for high spin fields. The resulting topologically massive gravity has been studied for almost thirty years since its proposal\cite{Deser:1982vy, Deser:1981wh}. It was found generically such a theory has an extra local massive mode, but it suffers from either the instability of the vacuum or negative energy of the black hole. However, few years ago, it was found that at a critical point, the theory could be well-defined, especially after imposing Brown-Henneaux boundary condition. At the critical point, the theory is called chiral gravity\cite{chiral,Maloney:2009ck}, as the local massive mode becomes degenerate with the left massless mode and the only degree of freedom is a massless boundary graviton. It was conjectured that chiral gravity is dual to a chiral CFT with only right-mover.
It is remarkable that even at the chiral point, there is actually a logarithmic mode, which carries negative energy\cite{CDWW1,CDWW2}. It has been suggested that there exists a logarithmic CFT dual to the TMG at the chiral point\cite{Grumiller:2008qz,arXiv:0906.4926,arXiv:1007.5189}, and the chiral gravity could only be a consistently truncated subsector of the whole theory after imposing the Brown-Henneaux boundary condition\cite{Maloney:2009ck}.

It would be interesting to investigate if there exist some kind of generalization of gravitational Chern-Simons term for the higher spin field and how the high spin field coupled to the topologically massive gravity. Actually, in \cite{Damour:1987vm}, a Chern-Simons-like term for spin-3 field has been proposed in flat spacetime. And very recently, the high spin topologically massive gravity has been studied. In \cite{Spin-3 TMG:2011}, the action of spin-3 topologically massive gravity has been
proposed in the first order formulation. The action could be rewritten as the sum of two Chern-Simons gauge field action with different levels, and   two extra terms imposing the torsion free conditions. The extra terms are essential to impose the torsion free condition and moreover leads to a massive mode for the fluctuations around the AdS$_3$ vacuum. The action not only includes the gravitational Chern-Simons term for the graviton, but also the Chern-Simons-like term for the spin-3 field in AdS$_3$\cite{Sahoo 1:2011}. In \cite{Spin-3 TMG:2011}, it was shown that at the critical point, the traceless part of spin-3 fluctuation has only massless boundary right-mover, keeping the chiral nature of the theory\footnote{This is actually true only when some generalized Brown-Henneaux boundary condition has been imposed such that the logarithmic mode carrying negative energy could be truncated consistently. The existence of the logarithmic modes at the critical points was found in both the traceless part\cite{Spin-3 TMG:2011} and the trace part\cite{Sahoo 1:2011}. }. The trace part was later investigated in \cite{Sahoo 1:2011}. However, in order to study the other high spin fields, the extra terms have to be introduced to impose the torsion free condition for every spin field. This incumber the study of higher spin topologically massive gravity. Another
disadvantage in using extra torsion-imposing term is that the action is not gauge invariant, since the Chern-Simons parts are written in terms of the gauge fields while the extra terms are linear in torsions.

In this paper, we propose an action for high spin topologically massive  gravity. In this new action, all of  the extra terms imposing the torsion free conditions for the high spin fields could be encoded into a single term, which is linearly proportional to the difference of the field strength of two gauge potential. Starting from this action, we discuss the gauge symmetry of the action and investigate the fluctuations around the AdS$_3$ vacuum. With the re-emergent linearized gauge symmetry in the background of AdS$_3$, we show  how to write vierbein as symmetric doubly traceless field and its trace. Moreover, we manage to study all the high spin fluctuations and obtain their equations of motion, which are third order differential equations. These equations reproduce exactly the existing results in the literature\cite{Sahoo 2:2011}. It is remarkable that in \cite{Sahoo 1:2011,Sahoo 2:2011}, the equations of motion of high spin fluctuations around AdS$_3$ were derived from the linearized action for high-spin Frondal fields with Chern-Simons-like terms, where the coefficients of the Chern-Simons-like terms for the spin higher than $3$ were set by demanding the chiral point of high spin fields coincide with the one of gravity. In this paper, our analysis is based on the first order action of the frame-like fields and spin-connections. We show how to relate the fluctuations of the frame-like fields to the Frondal fields, and derive the equations of motion of all higher spin fluctuations. Therefore the match of the equations of motion suggest that the Chern-Simons-like terms suggested in \cite{Sahoo 2:2011} are correct on one hand, and our theory is reasonable on the other hand. Certainly, our action includes not only the linearized action, but also the interactions among the graviton and other higher spin fields.

The remaining parts of the paper are organized as follows. In the next section, we introduce the action of high spin topologically massive  gravity. In Sec. 3, we discuss the symmetry of the action. In Sec. 4, we discuss the fluctuations in AdS$_3$ vacuum and obtain their equations of motion. We end with some discussions in Sec. 5. In the appendices, we include
the convention on $SL(n,R)$ algebra, and some technical details.

\section{Action}

 We start from the following action
\be\label{action}
S_{TMG}=\frac{k}{4\pi}[(1-\frac{1}{\mu l})S_{CS}[A]-(1+\frac{1}{\mu l})S_{CS}[\bar{A}]-\frac{1}{\mu}\int tr(\beta\wedge(F-\bar{F}))].
\ee
where we have the Chern-Simons action as
\be
S_{CS}[A]=\int tr(A\wedge dA+\frac{2}{3}A\wedge A\wedge A),
\ee
the gauge curvature
\be
F=dA+A\wedge A, \hs{3ex}\bar{F}=d\bar{A}+\bar{A}\wedge\bar{A}.
\ee
and one-form Lagrangian multiplier $\beta$. The gauge field $A$, $\bar{A}$ and the Lagrangian multiplier $\beta$ are in the adjoint representation of the corresponding group.  The gauge field $A$ and $\bar{A}$ are related to the frame-like fields and spin connection as\footnote{From now on we denote the $AdS$ radius $l=1$.}
\be
A=\omega+e,\hs{3ex}\bar{A}=\omega-e,
\ee

There are several remarks on the action (\ref{action}):
\begin{enumerate}
\item Firstly, the action is defined without explicit dependence on the gauge group. In other words, as an action on gauge fields, it make sense for any group which can be interpreted as higher spin field coupled to gravity. In this paper we consider the group $SL(n,R)\times SL(n,R)$ and discuss the fields of spin from 2 to $n$;
\item Secondly, when $\mu\to\infty$, it reduces to the action of the well-known high spin AdS$_3$ gravity which describes a tower of higher spin fields from spin 3 to spin n coupled
to gravity provided  the gauge group is chosen to be $SL(n,R)\times SL(n,R)$ \cite{Theisen:2010};
 \item Thirdly, when $\beta\not=0$, the last term is a Lagrangian multiplier.
The imposed condition $F=\bar{F}$ looks strange, but it is nothing but torsion-free condition.  If the gauge group is chosen
as $SL(2,R)\times SL(2,R)$, then the condition reduces to the torsion-free condition on the spin connection and the first-order action (\ref{action}) is equivalent to the famous topologically massive gravity\cite{Deser:1982vy, Deser:1981wh}. If the gauge group is chosen to be $SL(3,R)\times SL(3,R)$, the conditions once again give the torsion-free conditions on both spin-2 and spin-3 connection, and the action defines the so-called spin-3 topologically massive gravity\cite{Spin-3 TMG:2011};
\item For general gauge group $SL(n,R)\times SL(n,R)$, the condition $F=\bar{F}$ encodes all the torsion-free conditions on the spin-connections, where the torsions are the same as the ones in usual higher spin AdS$_3$ gravity, which could be defined along the line in \cite{Theisen:2010}. In this case, the action describe all the high spin fields coupled to topological massive gravity. In the action, there are not only topologically Chern-Simons term for graviton, but also the similar parity-breaking Chern-Simons terms for higher spin fields. The theory will be called by us as high spin topologically massive gravity;
\item Finally, the action is defined in terms of the gauge potential, without the torsion appearing explicitly. As a result, the equations of motion could be given in a concise form
\begin{eqnarray}
(1-\frac{1}{\mu })F-\frac{1}{2\mu}(d\beta+\beta\wedge A+A\wedge\beta)=0\\
(1+\frac{1}{\mu })\bar{F}-\frac{1}{2\mu}(d\beta+\beta\wedge \bar{A}+\bar{A}\wedge\beta)=0\\
F=\bar{F}.
\end{eqnarray}
These equations are easier to deal with. In particular when we investigate the fluctuations around the AdS$_3$ vacuum, the analysis is quite simple, even if we consider an arbitrary spin $n$ fluctuation.
\end{enumerate}

\section{Symmetry}

In this section we study the symmetry of the action which is essential for us to interpret it as a higher spin generalization of topologically massive gravity. Let us start from
the case $\beta=0$. In this case, the action is invariant under the gauge transformation
\be
\delta A=d\lambda+[A,\lambda], \hs{3ex}\delta\bar{A}=d\bar{\lambda}+[\bar{A},\bar\lambda].
\ee
Another way to represent this gauge transformation is to introduce $\L$ and $\tilde{\L}$ by the relation
\be
\lambda=\Lambda+\tilde\Lambda, \hs{3ex}\bar{\lambda}=\Lambda-\tilde\Lambda,
\ee
then
\be
\delta_{\Lambda} A=d\Lambda+[A,\Lambda],\hs{3ex}\delta_{\Lambda} \bar{A}=d\Lambda+[\bar{A},\Lambda]
\ee
and
\be
\delta_{\tilde{\Lambda}} A=d\tilde{\Lambda}+[A,\tilde{\Lambda}],\hs{3ex}\delta_{\tilde{\Lambda}} \bar{A}=-d\tilde{\Lambda}-[\bar{A},\tilde{\Lambda}].
\ee
$\d_\Lambda$ induces the usual Lorentz transformation while $\d_{\tilde\Lambda}$ is  related to the gauge transformations of physical fields.

Next we turn to the case $\beta\not=0$. We observe that in the action the term linear in $\b$ breaks the previous gauge symmetry since $F$ and $\bar{F}$ changes as
\be
F\to F+[F,\lambda], \hs{3ex}\bar{F}\to \bar{F}+[\bar{F},\bar{\lambda}].
\ee
Nevertheless, the Lorentz transformation is preserved. Actually, under the transformation
\be
\delta_{\Lambda} A=d\Lambda+[A,\Lambda],\hs{3ex}\delta_{\Lambda} \bar{A}=d\Lambda+[\bar{A},\Lambda],\hs{3ex}\delta_{\Lambda}\beta=[\beta,\Lambda]
\ee
the action is invariant. On the other hand, the gauge symmetry involving $\bar{\L}$ is broken. Namely, the imposition of the  torsion free condition breaks half of the gauge symmetries. This seems bad since the broken gauge symmetry is important
for us to define the gauge transformations of the physical fields. However,this is not a serious problem for three reasons. The first reason is that
the symmetry is broken only off-shell. The equation of motion tells us that $[F,\beta]=[\bar{F},\beta]=0$ hence the symmetry is preserved
on shell. The second reason is  there is nothing wrong to
break the symmetry off-shell by hand in field theory.  In other words, if we ignore the gauge symmetry which is missing in our action, we
just define a tower of higher rank field coupled to gravity. The third but most important reason is that although the half gauge symmetry is broken off-shell, if we focus on
the fluctuations around AdS$_3$ background, the broken gauge symmetry could be recovered at linearized level.

The emergence of the linearized gauge symmetry in the AdS$_3$ vacuum can be shown as follows. It is easy to see that the AdS$_3$ background with all higher spin fields vanishing is a solution of our
theory. We take it as the vacuum. Considering the fluctuations around the vacuum, we expand  the gauge fields as $A+a$ and $\bar{A}+\bar{a}$, where we denote the background
field as $A$ and $\bar{A}$. Since in the vacuum $\beta=0$, we use $\beta$ to denote the fluctuations in the following discussion. Then we expand the
original action in the AdS$_3$ vacuum to quadratic order of the fluctuations and have
\be
S_{(2)}=\frac{k}{4\pi}(S[a]-S[\bar{a}]+S_{\beta})
\ee
where
\be
S[a]=(1-\frac{1}{\mu })\int tr( (da+A\wedge a+a\wedge A)\wedge a),
\ee
\be
S[\bar{a}]=(1+\frac{1}{\mu})\int tr ((d\bar{a}+\bar{A}\wedge\bar{a}+\bar{a}\wedge\bar{A})\wedge\bar{a}),
\ee
\be
S_{\beta}=-\frac{1}{\mu}\int tr(\beta\wedge[(da+A\wedge a+a\wedge A)-(d\bar{a}+\bar{A}\wedge\bar{a}+\bar{a}\wedge\bar{A})]).
\ee
Note that it is important to take the fluctuations $\b$ into account as it induces
extra massive degrees of freedom, as we will show later.
We can show that the action is invariant under the following gauge transformations
\be
\delta_{\Lambda}a=d\Lambda+[A,\Lambda],\hs{3ex}\delta_{\Lambda}\bar{a}=d\Lambda+[\bar{A},\Lambda],\hs{3ex}\delta_{\Lambda}\beta=0
\ee
and
\be
\delta_{\tilde{\Lambda}}a=d\tilde{\Lambda}+[A,\tilde{\Lambda}],\hs{3ex}\delta_{\tilde{\Lambda}}\bar{a}=-d\tilde{\Lambda}-[\bar{A},\tilde{\Lambda}],
\hs{3ex}\delta_{\tilde{\Lambda}}\beta=0.
\ee
Note that the second gauge symmetry has been broken in the original action but it  re-emerges in the background AdS$_3$. This owes to the fact that
in the AdS$_3$ background $F=\bar{F}=0$.

Due to the re-emergence of the linearized gauge symmetry generated by $\tilde{\Lambda}$, we can define the spin-$s$ fluctuation gauge fields as
\be
h_{\nu, \nu_1\cdots \nu_{s-1}}=\bar{e}_{\nu_1}^{\ a_1}\cdots\bar{e}_{\nu_{s-1}}^{\ a_{s-1}}e_{\nu a_1\cdots a_{s-1}}
\ee
where $\bar{e}$ is $AdS_3$ background vierbein and we have chosen a general $SL(n,R)\times SL(n,R)$ bases following  \cite{Theisen:2010}. The explicit form of the bases is given in Appendix A. After a Lorentz transformation we can
set $h_{\nu,\nu_1\cdots\nu_{s-1}}$ field to a symmetric field
\be
\Phi_{\nu \nu_1\cdots \nu_{s-1}}=\frac{1}{s}\bar{e}_{(\nu_1}^{\ a_1}\cdots\bar{e}_{\nu_{s-1}}^{\ a_{s-1}}e_{\nu) a_1\cdots a_{s-1}}
\ee
which has a gauge
symmetry generated by $\xi_{\nu_1\cdots\nu_{s-1}}=\bar{e}_{\nu_1}^{\ a_1}\cdots\bar{e}_{\nu_{s-1}}^{\ a_{s-1}}\tilde{\Lambda}_{a_1\cdots a_{s-1}}$
\be
\delta_{\xi}\Phi_{\nu \nu_1\cdots \nu_{s-1}}=\nabla_{(\nu}\xi_{\nu_1\cdots\nu_{s-1})}.
\ee
From the definition of $\xi_{\nu_1\cdots\nu_{s-1}}$, we know that it is symmetric and traceless since $\tilde{\Lambda}$ is symmetric and traceless. Also, the traceless condition of
$e_{\nu}^{\ a_1\cdots a_{s-1}}$ induces the double traceless condition of $\Phi_{\nu\nu_1\cdots\nu_{s-1}}$.

Since the procedure of symmetrization from $h$ field to $\Phi$ field is a key point in our discussion of the fluctuations, we
would like to clarify this procedure more clearly. This has also been discussed in \cite{Vasiliev:1980as} in the flat spacetime background. Note that in our first order formulation, the Lorentz transformation becomes a gauge transformation generated
by $\Lambda$. When we go back to the physical fields of higher spin $s$, we need to choose a gauge to avoid this redundant degree of freedom.
$\Lambda$ can be parameterized as
\be
\Lambda=\sum_{s=2}^{s=n}\Lambda^{a_1\cdots a_{s-1}}T_{a_1\cdots a_{s-1}}
\ee
 where $\Lambda^{a_1\cdots a_{s-1}}$ is symmetric and traceless
and the $SL(n,R)$ generators $T_{a_1\cdots a_{s-1}}$ are defined in Appendix A. When one goes from $h$ to $\Phi$, it is necessary to choose a gauge to eliminate the redundant degree of freedom
generated by these $\Lambda^{a_1\cdots a_{s-1}}$. This is always possible since $h_{\nu, \nu_1\cdots \nu_{s-1}}$ has $3(2s-1)$ independent components while
$\Phi_{\nu \nu_1\cdots \nu_{s-1}}$ has $2(2s-1)$ components. Their difference  $(2s-1)$, is simply the number of independent components of symmetric traceless $\Lambda^{a_1\cdots a_{s-1}}$.
We can make a decomposition of $h$ schematically as
\be
h\sim\Phi+\Lambda.
\ee
The precise decomposition  depends on their special properties, which  we list them  as follows
\be
h_{(\nu,\nu_1\cdots\nu_{s-1})}=s\Phi_{\nu\nu_1\nu_{s-1}},\hs{3ex}h_{\nu,\sigma\nu_1\cdots\nu_{s-3}}^{\ \sigma}=0
\ee
\be
\Phi^{\nu\sigma}_{\   \nu\sigma\nu_1\cdots\nu_{s-4}}=0
\ee
\be
\Theta_{(\nu,\nu_1\cdots\nu_{s-1})}=0,\hs{3ex}\Theta^{\nu}_{\ \nu\nu_2\cdots\nu_{s-1}}=0,\hs{3ex}\Theta_{\nu, \sigma\nu_3\cdots\nu_{s-1}}^{\ \sigma}=0.
\ee
Here $\Theta$ is defined as
\be
\Theta_{\nu,\nu_1\cdots\nu_{s-1}}=\epsilon^{\kappa\sigma}_{\ \      (\nu_1}\bar{e}_{|\nu\kappa}\Lambda_{\sigma|\nu_2\cdots\nu_{s-1})},
\ee
and appears in the Lorentz transformation of $h_{\nu\nu_1\cdots\nu_{s-1}}$.
Then $h$ can be expressed as
\begin{eqnarray}
h_{\nu,\nu_1\cdots\nu_{s-1}}&=&\Phi_{\nu\nu_1\cdots\nu_{s-1}}+\Theta_{\nu,\nu_1\cdots\nu_{s-1}}+\frac{s-2}{2(s-1)}g_{\nu(\nu_1}\Phi'_{\nu_2\cdots \nu_{s-1})}
\nonumber\\& & -\frac{1}{s-1}g_{(\nu_1\nu_2}\Phi'_{|\nu|\nu_3\cdots\nu_{s-1})}.\label{decom}
\end{eqnarray}
where we use prime to denote taking the trace part of the corresponding fields. We also use the convention that the symmetrization symbol denotes the minimal number of terms it encloses and without any normalization. This expression satisfies all the properties of $h$ and $\Phi$. Note that the naive decomposition of $h=\Phi+\Lambda$ is wrong since it violates the
properties of $h$ and $\Phi$. In the next section, we will deal with the fluctuations around $AdS_3$ background and the decomposition (\ref{decom}) is important for
our discussion.

\section{Fluctuations around AdS$_3$ vacuum}

From the quadratical  action of the fluctuations around $AdS_3$ vacuum we can derive their equations of motion
\begin{eqnarray}
G(da+a\wedge A+A\wedge a)&=&d\beta+\beta\wedge A+A\wedge \beta\\
\bar{G}(d\bar{a}+\bar{a}\wedge\bar{A}+\bar{A}\wedge\bar{a})&=&d\beta+\beta\wedge \bar{A}+\bar{A}\wedge \beta\\
da+a\wedge A+A\wedge a&=&d\bar{a}+\bar{a}\wedge\bar{A}+\bar{A}\wedge\bar{a}
\end{eqnarray}
where we have defined $G=2(\mu-1),\bar{G}=2(\mu+1)$. We find that in the $AdS_3$ background, this linearized equations
describe free fluctuations of spin 2 to spin $n$ once we choose the basis of $SL(n,R)$ in the appendix A. The interaction between different spin is neglected
since they are higher order terms.

When $\beta=0$, we use the previous decomposition of $h_{\nu\nu_1\cdots\nu_{s-1}}$ and choose a gauge that
$\Theta_{\nu\nu_1\cdots\nu_{s-1}}=0$, then we can derive the Fronsdal equation\cite{Fronsdal:1978}
\begin{eqnarray}
\mathcal{F}_{\nu_1\cdots\nu_s}\equiv \Box\Phi_{\nu_1\cdots\nu_{s}}-\nabla_{(\nu_1|}\nabla^{\sigma}\Phi_{\sigma|\nu_2\cdots\nu_{s})}+\frac{1}{2}\nabla_{(\nu_1}\nabla_{\nu_2}\Phi'_{\nu_3\cdots\nu_s)}\nonumber
\\-(s^2-3s)\Phi_{\nu_1\cdots\nu_{s}}-2g_{(\nu_1\nu_2}\Phi'_{\nu_3\cdots\nu_s)}=0
\end{eqnarray}

When $\beta\not=0$, in principle we will
get a third order differential equation for each spin $s$ field. Here we give
the algorithm to derive this equation in our formulation. Note that this algorithm still holds when $\beta=0$.\\\\
Step 1: Decompose the veilbein fluctuations as
\begin{eqnarray}
h_{\nu,\nu_1\cdots\nu_{s-1}}&=&\Phi_{\nu\nu_1\cdots\nu_{s-1}}+\Theta_{\nu,\nu_1\cdots\nu_{s-1}}+\frac{s-2}{2(s-1)}g_{\nu(\nu_1}\Phi'_{\nu_2\cdots \nu_{s-1})}
\nonumber\\& & -\frac{1}{s-1}g_{(\nu_1\nu_2}\Phi'_{|\nu|\nu_3\cdots\nu_{s-1})}.
\end{eqnarray}
Step 2: Define a set of quantities as
\begin{eqnarray}
Q_{c, a_1\cdots a_{s-1}}=\epsilon_{c}^{\mu\nu}\nabla_{\mu}e_{\nu a_1\cdots a_{s-1}},\hs{3ex}
X_{c,a_1\cdots a_{s-1}}=Q_{c,a_1\cdots a_{s-1}}-\frac{1}{s}g_{c(a_1}Q'_{a_2\cdots a_{s-1})},\\
W_{c,a_1\cdots a_{s-1}}=\frac{1}{2}\epsilon_{c}^{\ \mu\nu}F_{\mu\nu a_1\cdots a_{s-1}},
\hs{3ex}
P_{c,a_1\cdots a_{s-1}}=W_{c,a_1\cdots a_{s-1}}-\frac{1}{s}g_{c(a_1}W'_{a_2\cdots a_{s-1})}.
\end{eqnarray}
Step 3: Use the equations of motion to derive the following relations
\begin{eqnarray}
\omega_{c,a_1\cdots a_{s-1}}&=&
X_{c,a_1\cdots a_{s-1}}-\frac{1}{s-1}X_{(c,a_1\cdots a_{s-1})},\\
\beta_{c,a_1\cdots a_{s-1}}&=&
2[P_{c,a_1\cdots a_{s-1}}-\frac{1}{s-1}P_{(c,a_1\cdots a_{s-1})}].
\end{eqnarray}
Step 4: Derive the equation of motion of $\beta$
\be
\epsilon_{c}^{\ \rho\nu}\nabla_{\rho}\beta_{\nu a_1\cdots a_{s-1}}=2\mu W_{c,a_1\cdots a_{s-1}}.
\ee
After symmetrization, one gets
\be
\epsilon_{(a_1|}^{\ \ \rho\nu}\nabla_{\rho}\beta_{\nu| a_2\cdots a_{s})}=2\mu W_{(a_1,a_2\cdots a_{s})}.
\ee
Step 5: Choose a gauge $\Theta_{\nu\nu_1\cdots\nu_{s-1}}=0$ and substitute $e_{\nu, a_1\cdots a_{s-1}}$ in terms of $\Phi_{\nu a_1\cdots a_{s-1}}$.

The calculation is quite tedious. The explicit form of $W_{c,a_1\cdots a_{s-1}}$ and $\beta_{c,a_1\cdots a_{s-1}}$ are shown in Appendix B. For arbitrary spin $s\ge 2$, we finally obtain the equations of the physical fields
\be
\mathcal{F}_{a_1\cdots a_s}+\frac{1}{\mu s(s-1)}\epsilon_{(a_1|}^{\ bc}\nabla_{b}\mathcal{F}_{c|a_2\cdots a_s)}=0.
\ee
To discuss free equations of motion of the fluctuations, we only need to know the commutation relation between SL(2,R) generators and high spin generators. This allows us to obtain the equations of motion for arbitrary spin up to $n$ in our formulation.

It is remarkable that the above equations are exactly those appear in \cite{Sahoo 1:2011,Sahoo 2:2011}. Note that in \cite{Sahoo 1:2011}, the form of the Chern-Simons-like term for spin-3 field was determined by the gauge invariance, with the coefficient being fixed by compare with the Chern-Simons gravity action. And in \cite{Sahoo 2:2011}, the higher spin generalization of such Chern-Simons-like term was proposed, by taking into account the gauge invariance and the chiral nature of the theory. Therefore, our result suggest that  this kind of term can be completely determined from our action.

\section{Discussion}

In this paper, we studied the high spin topologically massive gravity in the AdS$_3$ vacuum. We firstly proposed an action (\ref{action}) to describe the high spin fields coupled to the topologically gravity. One remarkable feature of the action is that the term imposing torsion free conditions is written in terms of gauge field strength. There is great advantage to use this action:
\begin{itemize}
\item It makes the  equations of motion easier to solve. One may just solve the equations of motions for the gauge fields and the Lagrangian multiplier, without
    paying special attention to the torsion free condition;
\item It makes the analysis of the fluctuations around the vacuum much easier. It allows us to analyze all the fluctuations from spin 2 up to spin $n$;
    \end{itemize}
We found that the imposition of the torsion free condition broke the gauge symmetry off-shell, but there is still on-shell gauge symmetry. Moreover, around the AdS$_3$ vacuum, the quadratic action of the fluctuations are gauge invariant. This allows us to decompose the fluctuations using the gauge symmetry.
We showed that in our theory, the spin $s$ field has a massive propagating mode for generic value of $\m l$, satisfying a third order differential equation.

In this paper, we focused on the gauge group $SL(n,R)\times SL(n,R)$ and obtained a  high spin topologically massive gravity with a negative cosmological constant. We may choose other gauge groups. One option is to consider a supergroup, which could lead to a supersymmetric theory. Another interesting option is to consider $SU(n)\times SU(n)$ group, which may describe a high spin topologically massive  gravity on a Euclideanised AdS$_3$.

We showed that in the AdS$_3$ vacuum, the theory is well-defined with recovered gauge symmetry at the quadratic order. There are other possible vacua in the theory. Even with the vanishing high spin fields, the gravity sector allows the warped spacetime, which has been
conjectured to be holographically dual to a conformal field theory\cite{Andy08}. In such warped spacetime, the gauge symmetry of the fluctuations needs to be investigated carefully. It is quite possible that the theory is still well-defined and thus provides a nontrivial example for high spin fields coupled to gravity beyond the maximally symmetry spaces. It would be very interesting to study these cases.

The action (\ref{action}) is more tractable than other forms of action in terms of the torsion. It may allows us to classify all the solutions. Even for the pure topologically
massive gravity, it is not clear if there exists asymptotical AdS$_3$ solution, which is
not an Einstein manifold. It would be interesting to address this issue
in our framework. Moreover, it would be important to search for other asymptotically AdS$_3$ solutions with nontrivial high spin hair.

\noindent
 {\large{\bf Acknowledgments}}

We are indebted to Jun-bao Wu  for many valuable discussions, and giving us a hardcopy of Ref. \cite{Vasiliev:1980as}, which is crucial for the accomplishment of this project. The work was in part supported by NSFC Grant No. 10975005.

\section*{Appendix A}
We choose the same convention as \cite{Theisen:2010}. The generators of $SL(n,R)$ are chosen such that
\be
T_a=J_a
\ee
form $SL(2,R)$ algebra
\be
[J_a,J_b]=\epsilon_{abc}J^c
\ee
and
\be
[J_a,T_{a_1\cdots a_{s-1}}]=\epsilon^{m}_{\ a(a_1}T_{a_2\cdots a_{s-1})m}
\ee
for $s>2$. The commutation relation between $T_{a_1\cdots a_{s-1}}$ for $s>2$ is irrelevant for our discussion.

\section*{Appendix B}
In this section, we show the explicit form of $W_{c,a_1\cdots a_{s-1}}$ and $\beta_{c,a_1\cdots a_{s-1}}$ in terms of the totally symmetric and doubly traceless field $\Phi_{a_1\cdots a_s}$.
\begin{eqnarray}
W_{c,a_1\cdots a_{s-1}}&=&-\frac{1}{s-1}\Box \Phi_{ca_1\cdots a_{s-1}}+\frac{s^2-3s}{s-1}\Phi_{ca_1\cdots a_{s-1}}+\frac{1}{s-1}\nabla_c\nabla^{\sigma}\Phi_{\sigma a_1\cdots a_{s-1}}\nonumber\\
& &+\frac{1}{s-1}\nabla_{(a_1|}\nabla^{\sigma}\Phi_{\sigma c|a_2\cdots a_{s-1})}-\frac{1}{s-1}g_{c(a_1|}\nabla^{\sigma}\nabla^{\nu}\Phi_{\sigma\nu|a_2\cdots a_{s-1})}\nonumber\\
& &-\frac{(s-2)(2s+1)}{2(s-1)}g_{c(a_1}\Phi'_{a_2\cdots a_{s-1})}+\frac{1}{s-1}g_{(a_1a_2|}\Phi'_{c|a_3\cdots a_{s-1})}\nonumber\\
& &+\frac{1}{2(s-1)}g_{c(a_1}\nabla_{a_2|}\nabla^{\sigma}\Phi'_{\sigma| a_3\cdots a_{s-1})}-\frac{1}{2(s-1)}\nabla_{(a_1}\nabla_{a_2|}\Phi'_{c|a_3\cdots a_{s-1})}\nonumber\\
& &+\frac{1}{s-1}g_{c(a_1}\Box\Phi'_{a_2\cdots a_{s-1})}-\frac{1}{s-1}\nabla_{(a_1|}\nabla_c\Phi'_{|a_2\cdots a_{s-1})}.
\end{eqnarray}
Note that when $\beta=0$, we find that $W_{c,a_1\cdots a_{s-1}}=0$. To get the Fronsdal equation, we need using this equation to eliminate the term related to $g_{c(a_1}\Box\Phi'_{a_2\cdots a_{s-1})}$ and then taking its symmetrization. The $\beta$ is
\begin{eqnarray}
&&\frac{(s-1)^2}{2}\beta_{c,a_1\cdots a_{s-1}}=\Box\Phi_{ca_1\cdots a_{s-1}}-(s^2-3s)\Phi_{ca_1\cdots a_{s-1}}-\nabla_c\nabla^{\sigma}\Phi_{\sigma a_1\cdots a_{s-1}}\nonumber\\
&&-\nabla_{(a_1|}\nabla^{\sigma}\Phi_{\sigma c|a_2\cdots a_{s-1})}-\frac{s-3}{s}g_{c(a_1|}\nabla^{\sigma}\nabla^{\rho}\Phi_{\sigma\rho|a_2\cdots a_{s-1})}
+\frac{2}{s}g_{(a_1a_2|}\nabla^{\sigma}\nabla^{\rho}\Phi_{\sigma\rho c|a_3\cdots a_{s-1})}\nonumber\\
&&+\frac{s-3}{s}g_{c(a_1}\Box\Phi'_{a_2\cdots a_{s-1})}
-\frac{2}{s}g_{(a_1a_2|}\Box\Phi'_{c|a_3\cdots a_{s-1})}-\frac{s^2-5s+8}{2}g_{c(a_1}\Phi'_{a_2\cdots a_{s-1})}\nonumber\\
&&+(s-3)g_{(a_1a_2|}\Phi'_{c|a_3\cdots a_{s-1})}+\frac{s}{2}\nabla_c\nabla_{(a_1}\Phi'_{a_2\cdots a_{s-1})}-\frac{s-2}{2}\nabla_{(a_1|}\nabla_c\Phi'_{|a_2\cdots a_{s-1})}\nonumber\\
&&+\frac{1}{2}\nabla_{(a_1}\nabla_{a_2}\Phi'_{a_3\cdots a_{s-1})c}
+\frac{s-3}{2s}g_{c(a_1}\nabla_{a_2|}\nabla^{\sigma}\Phi'_{\sigma|a_3\cdots a_{s-1})}\nonumber\\
&&-\frac{1}{s}g_{(a_1a_2|}\nabla_{c}\nabla^{\sigma}\Phi'_{\sigma|a_3\cdots a_{s-1})}-\frac{1}{s}g_{(a_1a_2}\nabla_{a_3|}\nabla^{\sigma}\Phi'_{\sigma c|a_4\cdots a_{s-1})}
\end{eqnarray}

\end{document}